\documentclass[12pt]{article}
\usepackage{amssymb}
\begin{document}
\begin{center}
In-medium QCD 
and Cherenkov gluons

\bigskip

I.M. Dremin\footnote{email: dremin@lpi.ru}\\ 

{\it Lebedev Physical Institute, Moscow, Russia}

\end{center}

%

%
%
%

%
\abstract{The equations of in-medium gluodynamics are proposed. Their 
classical lowest order solution is explicitly shown for a color charge
moving with constant speed. For nuclear permittivity larger than 1 it
describes emission of Cherenkov gluons resembling results of classical
electrodynamics. The choice of nuclear permittivity and Lorentz-invariance
of the problem are discussed. Effects induced by the transversely and
longitudinally moving (relative to the collision axis) partons at LHC
energies are described.
 
} 
%
%
\section{Introduction}
\label{intro}
The collective effects observed in ultrarelativistic heavy-ion collisions at 
SPS and RHIC \cite{st, fw, ph, ul} have supported the conjecture of quark-gluon 
plasma (QGP)
formed in these processes. The properties and evolution of this medium are
widely debated. At the simplest level it is assumed to consist of a set of
current quarks and gluons. It happens however that their interaction is quite
strong so that the notion of the strongly interacting quark-gluon plasma
(sQGP) has been introduced. Moreover, this substance reminds an ideal
liquid rather than a gas. Whether perturbative quantum chromodynamics (pQCD)
is applicable to the description of the excitation modes of this matter is
doubtful. Correspondingly, the popular theoretical approaches use either
classical solutions of in-vacuum QCD equations at the initial stage or 
hydrodynamics at the final stage of its evolution.

The collective excitation modes of the medium may however play a crucial role.
One of the ways to gain more knowledge about the excitation modes is to 
consider the propagation of relativistic partons through this matter.
Phenomenologically their impact would be described by the nuclear permittivity
of the matter corresponding to its response to passing partons. Namely this
approach is most successful for electrodynamical processes in matter.
Therefore it is reasonable to modify QCD equations by taking into account
collective properties of the quark-gluon medium. For the sake of simplicity
we consider here the gluodynamics only. The generalization to quarks is
straightforward.

The classical lowest order solution of these equations coincides with
Abelian electrodynamical results up to a trivial color factor. One of the most
spectacular of them is Cherenkov radiation and its properties. Now, Cherenkov
gluons take place of Cherenkov photons \cite{d1, ko}. Their emission in high 
energy hadronic collisions is described by the same formulae but with nuclear
permittivity in place of the usual one. It should be properly defined.
Actually, one considers them as quasiparticles, i.e. quanta of the medium
excitations with properties determined by the permittivity. The interplay
of medium properties and velocity of the particle is crucial for the 
radiation field.

Another important problem of this approach is related to the notion of the
rest system of the medium. The Lorentz invariance is lost if the permittivity
is introduced\footnote{In principle, this deficiency is cured by the 
relativistic generalization of the notion of permittivity (e.g., see 
\cite{llif}).}. Therefore one has to choose the proper coordinate system
where its definition is at work. While it is simple for macroscopic media in
electrodynamics, one should consider partons moving in different directions
with different energies in case of heavy-ion collisions. It has direct impact
on properties of emitted particles. The fast evolution of the medium and its 
short lifetime differ it from common electrodynamical examples.

All these problems are discussed in what follows.  


\section{Equations of in-medium gluodynamics}
\label{eqs}
At the beginning let us remind the classical in-vacuum Yang-Mills equations
\begin{equation}
\label{f.1}
D_{\mu}F^{\mu \nu }=J^{\nu },
\end{equation}
\begin{equation}
\label{1}
F^{\mu \nu }=\partial ^{\mu }A^{\nu }-\partial ^{\nu }A^{\mu }-
ig[A^{\mu },A^{\nu }],
\end{equation}
where $A^{\mu}=A_a^{\mu}T_a; \; A_a (A_a^0\equiv \Phi_a, {\bf A}_a)$ are the 
gauge field (scalar and vector) potentials, the color matrices $T_a$ satisfy
the relation $[T_a, T_b]=if_{abc}T_c$, $\; D_{\mu }=\partial _{\mu }-ig[A_{\mu }, \cdot], \;\; 
J^{\nu }(\rho, {\bf j})$ is a classical source current, $\hbar=c=1$ and the 
metric tensor is $g^{\mu \nu }$=diag(+,--,--,--).

In the covariant gauge $\partial _{\mu }A^{\mu }=0$ they are written as 
\begin{equation}
\label{f.2}
\square A^{\mu }=J^{\mu }+ig[A_{\nu }, \partial ^{\nu }A^{\mu }+F^{\nu \mu }],
\end{equation}
where $\square $ is the d'Alembertian operator. It was shown \cite{kovc} 
(and is confirmed in what follows) that in this gauge the classical gluon 
field is given by the solution of the corresponding Abelian problem.

The chromoelectric and chromomagnetic fields are
\begin{equation}
\label{2}
E^{\mu}=F^{\mu 0 },
\end{equation}
\begin{equation}
\label{3}
B^{\mu}=-\frac {1}{2}\epsilon ^{\mu ij}F^{ij},
\end{equation}
or as functions of gauge potentials in vector notations
\begin{equation}
\label{4}
{\bf E}_a=-{\rm grad }\Phi  _a-\frac {\partial {\bf A}_a}{\partial t}+
gf_{abc}{\bf A}_b \Phi _c,
\end{equation}
\begin{equation}
\label{5}
{\bf B}_a={\rm curl }{\bf A}_a-\frac {1}{2}gf_{abc}[{\bf A}_b{\bf A}_c].
\end{equation}
The equations of motion (\ref{f.1}) in vector form are written as
\begin{equation}
\label{6}
{\rm div } {\bf E}_a -gf_{abc}{\bf A}_b {\bf E}_c = \rho _a,
\end{equation}
\begin{equation}
\label{7}
{\rm curl } {\bf B}_a-\frac {\partial {\bf E}_a}{\partial t} - gf_{abc}
(\Phi _b {\bf E}_c+[{\bf A}_b {\bf B}_c])= {\bf j}_a.
\end{equation}
                                                                  
The Abelian equations of in-vacuum electrodynamics are obtained from 
Eq. (\ref{f.2}) if the second term in its right-hand side is put equal to 
zero and color indices omitted. The medium is accounted if $\bf E$ is
replaced by ${\bf D} =\epsilon {\bf E}$ in $F^{\mu \nu} \;$, i.e. in Eq.
(\ref{2})\footnote{$\epsilon $ denotes the dielectric permittivity of the
medium. The magnetic permittivity is put equal to 1 to simplify 
the formulae.}. Therefore the Eqs. (\ref{6}), (\ref{7}) in vector form are
most suitable for their generalization to in-medium case. 
The equations of in-medium electrodynamics differ from in-vacuum ones by
dielectric permittivity $\epsilon \neq 1$ entering there as
\begin{equation}
\label{f.3}
\bigtriangleup {\bf A}-\epsilon \frac{\partial ^2{\bf A}}{\partial t^2}=
-{\bf j},
\end{equation}
\begin{equation}
\label{f.4}
\epsilon (\bigtriangleup \Phi-\epsilon \frac{\partial ^2 \Phi}{\partial t^2})=
-\rho .
\end{equation}
The permittivity describes the matter response to the induced fields which is 
assumed to be linear and constant in Eqs. (\ref{f.3}), (\ref{f.4}). It is 
determined by the distribution of internal current sources in the medium.
Then external currents are only left in the right-hand sides of these equations. 

Now, the Lorentz gauge condition is
\begin{equation}
\label{f.5}
{\rm div } {\bf A}+\epsilon \frac {\partial \Phi }{\partial t}=0.
\end{equation}

The Lorentz invariance is broken if $\epsilon \neq 1$ in front of the second
terms in the left-hand sides. Then one has to deal within the coordinate system
where a substance is at rest. The values of $\epsilon $ are determined just
there. To cancel these requirements one must use Minkowski relations
between ${\bf {D, \; E, \; B, \; H}}$ valid for a moving medium \cite{llif}. 
It leads to more complicated formulae, and we do not use them in this paper.

The most important property of solutions of these equations is that
while the in-vacuum ($\epsilon = 1$) equations do not admit any radiation
processes, it happens for $\epsilon \neq 1$ that there are solutions of
these equations with non-zero Poynting vector.

Now we are ready to write down the equations of in-medium gluodynamics
generalizing Eq. (\ref{f.2}) in the same way as Eqs. (\ref{f.3}),
(\ref{f.4}) are derived in electrodynamics. We introduce the nuclear
permittivity and denote it also by $\epsilon $ since it will not lead
to any confusion. After that one should replace ${\bf E}_a$ in Eqs. (\ref{6}),
(\ref{7}) by $\epsilon {\bf E}_a$ and get:
\begin{equation}
\label{8}
\epsilon ({\rm div } {\bf E}_a-gf_{abc}{\bf A}_b {\bf E}_c)=\rho _a,
\end{equation}
\begin{equation}
\label{9}
{\rm curl } {\bf B}_a-\epsilon \frac {\partial {\bf E}_a}{\partial t} -
gf_{abc}(\epsilon \Phi _b{\bf E}_c + [{\bf A}_b{\bf B}_c])={\bf j}_a.
\end{equation}
The space-time dispersion of $\epsilon $ is neglected here.
 
In terms of potentials these equations are cast in the form:
\begin{eqnarray}
\bigtriangleup {\bf A}_a-\epsilon \frac{\partial ^2{\bf A}_a}{\partial t^2}=
-{\bf j}_a -
gf_{abc}(\frac {1}{2} ( {\rm curl } [{\bf A}_b, {\bf A}_c]+
[{\bf A}_b {\rm curl } {\bf A}_c])+\frac {\partial }
{\partial t}({\bf A}_b\Phi _c)-  \nonumber \\
\epsilon \Phi _b\frac 
{\partial {\bf A}_c}{\partial t}- 
\epsilon \Phi _b {\rm grad } \Phi _c-\frac {1}{2} gf_{cmn}
[{\bf A}_b[{\bf A}_m{\bf A}_n]]+g\epsilon f_{cmn}\Phi _b{\bf A}_m\Phi _n), \hfill \label{f.6}
\end{eqnarray}

\begin{eqnarray}
\bigtriangleup \Phi _a-\epsilon \frac {\partial ^2 \Phi _a}
{\partial t^2}=-\frac {\rho _a}{\epsilon }+ 
gf_{abc}(2{\bf A}_b {\rm grad }\Phi _c+{\bf A}_b
\frac {\partial {\bf A}_c}{\partial t}-\epsilon 
\frac {\partial \Phi _b}{\partial t}
\Phi _c)+  \nonumber  \\
g^2 f_{amn} f_{nlb} {\bf A}_m {\bf A}_l \Phi _b. \hfill  \label{f.7}
\end{eqnarray}
If the terms with explicitly shown coupling constant $g$ are omitted, one gets
the set of Abelian equations which differ from electrodynamical equations
(\ref{f.3}), (\ref{f.4}) by the color index $a$ only. Their solutions are 
shown in the next section. The external current is ascribed to a parton fast 
moving relative to other partons "at rest".

The potentials are linear in $g$ because the classical current $J^{\mu }$ is
linear also. Therefore omitted terms are of the order of $g^3$ and can be 
taken into account as a perturbation. It was done in \cite{kov, ma} for
in-vacuum gluodynamics. Here, the general procedure is the same. After
getting explicit lowest order solution (see the next section) one exploits
it together with the non-Abelian current conservation condition to find the
current component proportional to $g^3$. Then with the help of Eqs.
(\ref{f.6}), (\ref{f.7}) one finds the potentials up to the order $g^3$.
They can be represented as integrals convoluting the current with the
corresponding in-medium Green function. The higher order corrections may 
be obtained in the same way. We postpone their consideration for further
publications.

The crucial distinction between Eq. (\ref{f.2}) and Eqs. (\ref{f.6}),
(\ref{f.7}) is that there is no radiation (the field strength is zero in
the forward light-cone and no gluons are produced) in the lowest order solution 
of Eq. (\ref{f.2}) and it is admitted for Eqs. (\ref{f.6}), (\ref{f.7})
because $\epsilon $ takes into account the collective response (polarization)
of the nuclear matter. We have assumed that no color indices are attached to
$\epsilon $. It would correspond to the collective response of the 
color-neutral (on the average) medium if color exchange between the external
current $J^{\mu }$ and medium excitations is numerous and averages to zero.
The lack of knowledge about the collective excitations of the nuclear medium
prevents more detailed studies. However it seems to be justified at least
for Cherenkov effects.

\section{Cherenkov gluons as the classical lowest order solution of 
in-medium gluodynamics}
\label{CG}
Cherenkov effects are especially suited for treating them by classical
approach to Eqs. (\ref{f.6}), (\ref{f.7}). Their unique feature is
independence of the coherence of subsequent emissions on the time interval
between these processes. 

The problem of the coherence length for Cherenkov radiation was extensively 
studied \cite{tf, fr}. It was shown that the $\omega $-component of the field
of a current can be imitated by a set of oscillators with frequency $\omega $
situated along the trajectory. The waves from all oscillators add up in the
direction given by the Cherenkov angle $\theta $ independent on the length
of the interval filled in by these oscillators.
The phase disbalance $\Delta \phi $ between
emissions with frequency $\omega =k/\sqrt {\epsilon }$ separated by the
time interval $\Delta t $ (or the length $\Delta z=v\Delta t$) is given by
\begin{equation}
\label{f.9}
\Delta \phi =\omega \Delta t-k\Delta z\cos \theta =
k\Delta z(\frac {1}{v\sqrt {\epsilon }}-\cos \theta )
\end{equation}
up to terms which vanish for large distances between oscillating sources and
the detector. For Cherenkov effects the angle $\theta $ is
\begin{equation}
\label{f.10}
\cos \theta = \frac {1}{v\sqrt {\epsilon }}.
\end{equation}
The coherence condition $\Delta \phi =0$ is valid independent of $\Delta z $.
This is a crucial property specific for Cherenkov radiation only\footnote
{The requirement for $\Delta \phi $ to be a multiple of $2\pi $ (or a weaker
condition of being less or of the order of 1) in cases when Cherenkov condition 
is not satisfied imposes limits on the effective radiation length as it happens,
e.g., for Landau-Pomeranchuk or Ter-Mikaelyan effects.}. Thus the change of
color at emission vertices is not important if one considers a particular
$a$-th component of color fields produced at Cherenkov angle. Therefore the 
fields $(\Phi _a, {\bf A}_a)$ and the classical current for in-medium 
gluodynamics can be represented by the product of their electrodynamical 
expressions $(\Phi , {\bf A})$ and the color matrix $T_a$. As a result, one 
can neglect the "rotation" of color at emission vertices and use in the lowest 
order for Cherenkov gluons the well known formulae for Cherenkov photons just 
replacing $\alpha $ by $\alpha _SC_A$ for gluon currents in probabilities of 
their emission. Surely, there is radiation at angles different from the 
Cherenkov angle (\ref{f.10}). For such gluons one should take into account 
the coherence length and color rotation considering corresponding Wilson 
lines \cite{kwi}.

Let us remind the explicit Abelian solution for the current with velocity 
${\bf v}$ along $z$-axis
\begin{equation}
\label{f.11}
{\bf j}({\bf r},t)={\bf v}\rho ({\bf r},t)=4\pi g{\bf v}\delta({\bf r}-{\bf v}t).
\end{equation}

In the lowest order the solutions for scalar and vector potentials are
related so that
\begin{equation}
{\bf A}^{(1)}({\bf r},t)=\epsilon {\bf v} \Phi ^{(1)}({\bf r},t),   \hfill  \label{f.13}
\end{equation}
where the superscript (1) indicates the solutions of order $g$.

Therefore the explicit expressions for $\Phi $ suffice.
Using the Fourier transform, the lowest order solution of Eq. (\ref{f.4}) 
with account of (\ref{f.11}) can be cast in the form
\begin{equation}
\label{four}
\Phi ^{(1)}({\bf r},t)=\frac {g}{2\pi ^2\epsilon }\int d^3k
\frac {{\rm exp }[i{\bf k(r-v}t)]}{k^2-\epsilon ({\bf kv})^2}
\end{equation}

The integration over the angle in cylindrical coordinates gives the Bessel
function $J_0(k_{\perp}r_{\perp})$. Integrating over the longitudinal 
component $k_z$ with account of the poles due to the denominator\footnote{These
poles are at work only for Cherenkov radiation!} and then over 
the transverse one $k_{\perp}$, one gets the following expression for the scalar 
potential \cite{ru}
\begin{equation}
\label{f.12}
\Phi ^{(1)}({\bf r},t)=\frac {2g}{\epsilon }\frac {\theta
(vt-z-r_{\perp }\sqrt {\epsilon v^2-1})}{\sqrt {(vt-z)^2-r_{\perp} ^2
(\epsilon v^2-1)}},
\end{equation}

Here $r_{\perp }=\sqrt {x^2+y^2}$ is the cylindrical coordinate, $z$ is the
symmetry axis. The cone
\begin{equation}
\label{f.14}
z=vt-r_{\perp }\sqrt {\epsilon v^2-1}
\end{equation}
determines the position of the shock wave due to the $\theta $-function
in Eq. (\ref{f.12}). The field is localized within this cone. The Descartes
components of the Poynting vector are related according to Eqs. (\ref{f.12}),
(\ref{f.13}) by the formulae
\begin{equation}
\label{f.15}
S_x=-S_z\frac {(z-vt)x}{r_{\perp }^2}, \;\;\; 
S_y=-S_z\frac {(z-vt)y}{r_{\perp }^2},
\end{equation}
so that the direction of emitted gluons is perpendicular to the cone
(\ref{f.14}) and defined by the Cherenkov angle
\begin{equation}
\label{f.16}
\tan ^2\theta=\frac {S_x^2+S_y^2}{S_z^2}=\epsilon v^2-1,
\end{equation}
which coincides with (\ref{f.10}).

The higher order terms ($g^3$ ...) can be calculated using Eqs. (\ref{f.6}),
(\ref{f.7}).

The expression for the intensity of the radiation is given by the Tamm-Frank
formula (up to Casimir operators)
\begin{equation}
\label{f.17}
\frac {dE}{dl}=4\pi \alpha_S\int \omega d\omega (1-\frac {1}{v^2\epsilon}).
\end{equation}
It is well known that it leads to infinity for constant $\epsilon $.
The $\omega $-dependence of $\epsilon $ (its dispersion) usually solves the
problem. For absorbing media $\epsilon $ acquires the imaginary part.
The sharp front edge of the shock wave is smoothed.  The angular distribution 
of Cherenkov radiation widens. The $\delta $-function at the angle 
(\ref{f.10}), (\ref{f.17}) is replaced by the Breit-Wigner shape \cite {gr} 
with maximum at the same angle (but $\vert \epsilon \vert $ in place of 
$\epsilon $) and the width proportional to the imaginary part. 
Without absorption, the potential (\ref{f.12}) is infinite on the cone.
With absorption, it is finite everywhere except the cone vertex and is inverse
proportional to the distance from the vertex. For low absorption, the field on 
the cone increases as (${\rm Im }\, \epsilon )^{-1/2}$ (see \cite{dkl}).
Absorption induces also longitudinal excitations (chromoplasmons) which are
proportional to the imaginary part of $\epsilon $ and usually small compared
to transverse excitations. The magnetic permittivity is easily taken into 
account replacing $\epsilon $ by $\epsilon \mu $ in the Breit-Wigner formula.

In electrodynamics the permittivity of real substances depends on
$\omega $. Moreover it has the imaginary part determining the absorption.
E.g., ${\rm Re }\, \epsilon $ for water (see \cite{ja}) is approximately constant
in the visible light region ($\sqrt {\epsilon }\approx 1.34$), increases at
low $\omega $ and becomes smaller than 1 at high energies tending to 1
asymptotically. The absorption (${\rm Im }\, \epsilon $) is very small for 
visible light but dramatically icreases in nearby regions both at low and 
high frequencies. Theoretically this behavior is ascribed to various collective
excitations in the water relevant to its response to radiation with different 
frequencies. Among them the resonance excitations are quite prominent (see, 
e.g., \cite{fe}). Even in electrodynamics, the quantitative theory of this 
behavior is still lacking, however.

Then, what can we say about the nuclear permittivity?
 
\section{The nuclear permittivity}
\label{NP}
The partons constituting high energy hadrons or nuclei interact during 
the collision for a very short time. Nevertheless, there are experimental 
indications that an intermediate state of matter (CGC, QGP, nuclear fluid ...)
is formed and evolves. Those are J/$\psi $-suppression, jet quenching, 
collective flow ($v_2$),
Cherenkov rings of hadrons etc. They show that there is collective response of 
the nuclear matter to color currents moving in it. Unfortunately, our
knowledge of its internal excitation modes is very scarce, much smaller than 
in electrodynamics. 

The attempts to calculate the nuclear permittivity from first principles are
not very convincing. It can be obtained from the polarization operator.
The corresponding dispersion branches have been computed in the lowest order
perturbation theory \cite{kk, kl, we}. Then the properties of collective 
excitations have been studied in the framework of the thermal field
theories (for review see, e.g., \cite{bi}). Their results with additional 
phenomenological ad hoc assumption about the role of resonances were used
in a simplified model of scalar fields \cite{ko} to show that the nuclear 
permittivity can be larger than 1 that admits Cherenkov gluons.

Let us stress the difference between these approaches and our consideration.
In Refs. \cite{kk, kl, we, bi} the medium response to the {\it induced}
current is analyzed. Namely it determines the nuclear permittivity.
The permittivity is the internal property of a medium. Its quantitative 
description poses problems even in QED. It becomes more difficult task in
QCD where confinement is not understood. Therefore we did not yet attempt 
to compute the nuclear permittivity and introduced it purely
phenomenologically in analogy to in-medium electrodynamics. Our main goal
is to study the medium response to the {\it external} color current.
Cherenkov effect is proportional to $g^2$ according to Eq. (\ref{f.12}) if
$\epsilon $ is constant or chosen purely phenomenologically. However 
$\epsilon $ should tend to 1 at small $g$ and Cherenkov effect disappears.
Thus it is of the order of $g^4$ at small $g$. Mach waves in hydrodynamics 
\cite{mu} are of the same order. When the current $J^{(3)}$ is treated as 
external one in equations of in-vacuum gluodynamics \cite{kovc, kov, ma} 
the effect is proportional to $g^6$.

We prefer to use the general formulae of the scattering theory \cite{go} to 
estimate the nuclear permittivity. It is related to the refractive index $n$ 
of the medium:
\begin{equation}
\label{f.18}
\epsilon =n^2
\end{equation}
and the latter one is expressed \cite{go} through the real part of the forward 
scattering amplitude of refracted quanta\footnote{In electrodynamics these 
quanta are photons. In QCD those are gluons.} ${\rm Re}F(0^o,E)$ as
\begin{equation}
\label{f.19}
{\rm Re} n(E )=1+\Delta n_R =1+\frac {6m_{\pi }^3\nu }{E^2}{\rm Re }F(E) =
1+\frac {3m_{\pi }^3\nu }{4\pi E } \sigma (E )\rho (E ).  
\end{equation}
Here $E$ denotes the energy, $\nu $ is the number of scatterers within a single 
nucleon, $m_{\pi }$ the pion mass, $\sigma (E)$ the cross section and $\rho (E)$ 
the ratio of real to imaginary parts of the forward scattering amplitude $F(E)$. 
Thus the emission of Cherenkov gluons is possible only for processes with 
positive ${\rm Re} F(E)$ or $\rho (E)$. Unfortunately, we are unable to 
calculate directly in QCD these characteristics of gluons\footnote{We can only 
say that ${\rm Re} F(E)\propto g^2$ at small $g$ that confirms above estimates.} 
and have to rely
on analogies and our knowledge of properties of hadrons. The only experimental
facts we get about this medium are brought by particles registered at the final 
stage. They have some features in common which (one can hope!) are also
relevant for gluons as the carriers of the strong forces. Those are the resonant
behavior of amplitudes at rather low energies and positive real part of the
forward scattering amplitudes at very high energies for hadron-hadron and
photon-hadron processes as measured from the interference of the Coulomb and
hadronic parts of the amplitudes. ${\rm Re} F(0^o,E)$ is always positive 
(i.e., $n>1$) within the low-mass wings of the Breit-Wigner resonances.  
This shows that the necessary condition for Cherenkov effects $n>1$ is
satisfied at least within these two energy intervals. This fact was used
to describe experimental observations at SPS, RHIC and cosmic ray energies. 
The asymmetry of
the $\rho $-meson shape at SPS \cite{da} and azimuthal correlations of 
in-medium jets at RHIC \cite{ul, ajit} were explained by emission of 
comparatively low-energy Cherenkov gluons \cite{dnec, drem1}.
The parton density and intensity of the radiation were estimated. In its turn,
cosmic ray data \cite{apan} at energies corresponding to LHC ask for very high 
energy gluons to be emitted by the ultrarelativistic partons moving along the
collision axis \cite{d1}. Let us note the important difference from
electrodynamics where $n<1$ at high frequencies. For QGP the high-energy
condition $n>1$ is a consequence of its instability.

The dispersion ($\omega $-dependence of $n$) was taken into account. Otherwise
the intensity of the radiation given by Eq. (\ref{f.17}) diverges.  It can be 
easily incorporated in Eqs. (\ref{f.6}), (\ref{f.7}) (more precisely, in their
Fourier components). The formula (\ref{f.19}) valid for $n-1\ll 1$ is 
generalized to Lorenz-Lorentz expression for larger $n$. The imaginary part
of $\epsilon $ can be easily accounted. 
In principle, it may be estimated from RHIC data (see \cite{dkl}).

Up to now we did not discuss one of the most important problems of the
coordinate system in which the permittivity is defined.

\section{The rest system of the nuclear matter}
The in-medium equations are not Lorentz-invariant. There is no problem in
macroscopic electrodynamics because the rest system of the macroscopic matter
is well defined and its permittivity is considered there. For collisions of
two nuclei (or hadrons) it asks for special discussion.

Let us consider a particular parton which radiates in the nuclear matter.
It would "feel" the surrounding medium at rest if momenta of all other partons 
(or constituents of the matter), with which this parton can interact, 
sum to zero. In RHIC 
experiments the triggers which registered the jets (created by partons) 
were positioned at 90$^o$ to the collision axis. Such partons should be
produced by two initial forward-backward moving partons scattered at 90$^o$.
The total momentum of other partons (medium spectators) is balanced because 
for such geometry the partons from both nuclei play a role of spectators 
forming the medium. Thus the center
of mass system is the proper one to consider the nuclear matter at rest in
this experiment. The permittivity must be defined there. The Cherenkov rings 
consisting of hadrons have been registered around the away-side jet which
traversed the nuclear medium. This geometry requires however high statistics
because the rare process of scattering at 90$^o$ has been chosen. 

The forward 
(backward) moving partons are much more numerous and have higher energies. 
However, one can not treat the radiation of such a primary parton in c.m.s.
in the similar way because the momentum of spectators is different from zero
i.e. the matter is not at rest. Now the spectators (the medium) are formed
from the partons of another nucleus only. Then the rest system of the medium
coincides with the rest system of that nucleus and the permittivity should 
refer to this system. The Cherenkov radiation of such highly energetic 
partons must be considered there. That is what was done for interpretation
of the cosmic ray event in \cite{d1}. This discussion clearly shows that 
one must carefully define the rest system for other geometries of the 
experiment with triggers positioned at different angles.

Thus our conclusion is that the definition of $\epsilon $ depends on the 
experiment geometry. Its corollary is that partons moving in different
directions with different energies can "feel" different states of matter in
the ${\bf same}$ collision of two nuclei because of the dispersive dependence
of the permittivity. The transversely scattered partons with comparatively 
low energies can analyze the matter with rather large permittivity corresponding
to the resonance region while the forward moving partons with high energies 
would "observe" low permittivity in the same collision. This peculiar feature
can help scan the $(\ln x, Q^2)$-plane as it is discussed in \cite {drem2}.
It explains also the different values of $\epsilon $ needed for description
of RHIC and cosmic ray data.

 These conclusions can be checked at LHC because 
both RHIC and cosmic ray geometry will become available there. The energy
of the forward moving partons would exceed the thresholds above which 
$n>1$. Then both types of experiments can be done, i.e. the 90$^o$-trigger
and non-trigger forward-backward partons experiments. The predicted results for 
90$^o$-trigger geometry are similar to those at RHIC. The non-trigger
Cherenkov gluons should be emitted within the rings at polar angles of tens
degrees in c.m.s. at LHC by the forward moving partons (and symmetrically by 
the backward ones). This idea is supported by some events observed in cosmic
rays \cite{apan, drem1}.

\section{Conclusions}
\label{CONCL}
The equations of in-medium gluodynamics (\ref{f.6}), (\ref{f.7}) are proposed. 
They remind the
in-medium Maxwell equations with non-Abelian terms added. Their lowest
order classical solutions are similar (up to the trivial color factors) 
to those of electrodynamics (\ref{f.12}), especially, for Cherenkov gluons. 
The nuclear permittivity of the hadronic medium is related to the 
forward scattering hadronic amplitudes and its possible generalization 
is discussed. This definition asks for the distinction between 
the different coordinate systems in which the Cherenkov radiation
(and nuclear permittivity) should be treated for partons moving in 
different directions with different energies.

This consideration has led to explanation of several effects observed at
SPS, RHIC, cosmic ray energies and predicts new features at LHC \cite{drem2}. 
Some estimates of properties of the nuclear matter formed in ultrarelativistic
heavy-ion collisions have been done and are predicted.

\section*{Acknowledgements}
\vspace{0.5mm}
I thank  A.V. Leonidov for useful discussions. This work was supported
in parts by the RFBR grants 06-02-17051-a, 06-02-16864-a, 08-02-91000-CERN-a.

\end{document}